\documentclass[aps,cite,a4paper]{article}
\usepackage{epsfig}
\begin{document}
\begin{center}{\Large  \bf
Electronic response and bandstructure modulation of carbon nanotubes in a transverse electrical field}\\
\vspace{.5cm}
{Yan Li, Slava V. Rotkin, Umberto Ravaioli}\\
\vspace{.5cm}
{\it
  Beckman Institute for Advanced Science and Technology\\
  University of Illinois at Urbana-Champaign\\
  Urbana, IL 61801 USA}
\end{center}

\begin{abstract}
The electronic properties of carbon nanotubes (NTs) in a uniform transverse field are
investigated within a single orbital tight-binding (TB) model. For doped
nanotubes, the dielectric function is found to depend not only on
symmetry of the tube, but also on radius and Fermi level
position. Bandgap opening/closing is predicted for zigzag tubes, while it is 
found that armchair tubes always remain metallic, which is explained by the 
symmetry in their configuration. The bandstructures for both types are considerably 
modified when the field strength is large enough to mix neighboring subbands.
\end{abstract}

\section{Introduction}\label{sec:introduction}

The response of nanotubes to external electric fields is of interest both 
for transport devices\cite{transport} and for nano--electromechanical systems 
\cite{NEMS_exp,NEMS_theo}. When the transverse field strength is large enough to 
couple neighboring subbands, it is not appropriate to use the conventional 1D 
approximation, which assumes a uniform charge and potential distribution 
along the NT circumference. In this article, we address the issue of how the NT 
material properties are modified when the electronic potential 
along the circumference is no longer uniform.  We use a single $\pi$ orbital
tight-binding model in a self-consistent way to account for screening
effect caused by the non-uniform charge distribution along the NT circumference.

To our knowledge, this is the first study that considers a wide range of applied 
transverse electric field conditions.  Within the limit of relatively weak electric fields, 
we obtain different behaviors for semiconducting and  metallic
nanotubes, either intrinsic or doped. The weak field condition can be 
expressed as ${\cal E}\ll ta/(eR^2)$($\sim 0.1\mbox{V/\AA}$ for $R=8\mbox{\AA}$), where  
$t=2.5\mbox{eV}$ and $a=2.49\mbox{\AA}$ are the hopping 
integral and lattice constant of two-dimensional graphite
respectively, as defined for example in the book \cite{Dresselhaus}. One may safely use a rigid bond approximation in this
field limit, without severely perturbing the original bandstructure. 
We also investigate the regime of relatively strong fields, 
$ta/(eR^2)<{\cal E}\ll t/(ea)$, 
where bandstructure modifications need to be considered and subband
mixing cannot be neglected, while a TB approach can still be used.  
Also for this regime, qualitatively different behaviors of
semiconducting and metallic NTs are found.
 
The paper is organized as follows.
In Sec.\ref{sec:polarizability} we calculate the transverse polarizability 
for nanotubes of various symmetry, considering a linear approximation and no phonon 
contributions.  The dependence of the polarizability
on tube radius, symmetry and Fermi level position are discussed.  The role of the Fermi level
is not related to intra-band free carrier transitions, but rather
to inter-band transitions between neighboring conduction/valence bands. We then
exam the effect of a uniform transverse transverse field on
the bandstructure of nanotubes in Sec.\ref{sec:bandstructure}. 
Because of the symmetry of the nanotube, the subbands of the tube are mixed
according to certain selection rules and the bandstructure and bandgap are correspondingly
modified. Several interesting phenomena are predicted to happen in the
transverse field: bandgap opening/closing, energy subband flattening and
over-bending, lifting of subband
degeneracy, and generation of multiple valleys. These effects can be
used to modulate the bandgap, effective mass and carrier densities of
NTs and to enhance the density of states(DOS) at the Fermi level. For
example, it has been suggested that opening of
energy gaps of metalic nanotubes under a local transverse electric field
can be used for designing quantum switches\cite{Chang}. Our
calculation shows that there exists a critical magnitude of the field strength
beyond which the gap starts to decrease and oscillates further on.

For simplicity, only zigzag (both metallic and semiconducting)
and armchair tubes are considered in the full-band TB
calculations. The extension of the numerical results
to chiral tubes is confirmed by the analytical results within the
$\vec{k}\cdot\vec{p}$ method.

\section{Transverse polarizability of a NT in a weak electric field}\label{sec:polarizability}

An external electrical field creates an induced dipole moment $\emph{p}$ on the nanotube, which can be
estimated through the calculation of the polarizability of $\pi$
electrons \cite{Louie}. The unscreened linear polarizability $\alpha_0(\omega)$ relates the dipole moment
$p$ to the total electrical field ${\cal E}_{tot}$ with
$p=\alpha_0{\cal E}_{tot}$, which accounts for the contribution from the
single particle excitation. The dielectric function, defined as 
$\epsilon={\cal E}_{tot}/{\cal E}_{ex}$, and the actual (or screened) polarizability
$\alpha$ can be easily retrieved from $\alpha_0$ as

\begin{equation}
  \epsilon(\omega)=1+2\frac{\alpha_0(\omega)}{R^2}, \quad \alpha(\omega)=\frac{\alpha_0(\omega)}{\epsilon(\omega)}
\end{equation}

\begin{figure}[htb]
\centerline{
\includegraphics[angle=-90, width=5 in]{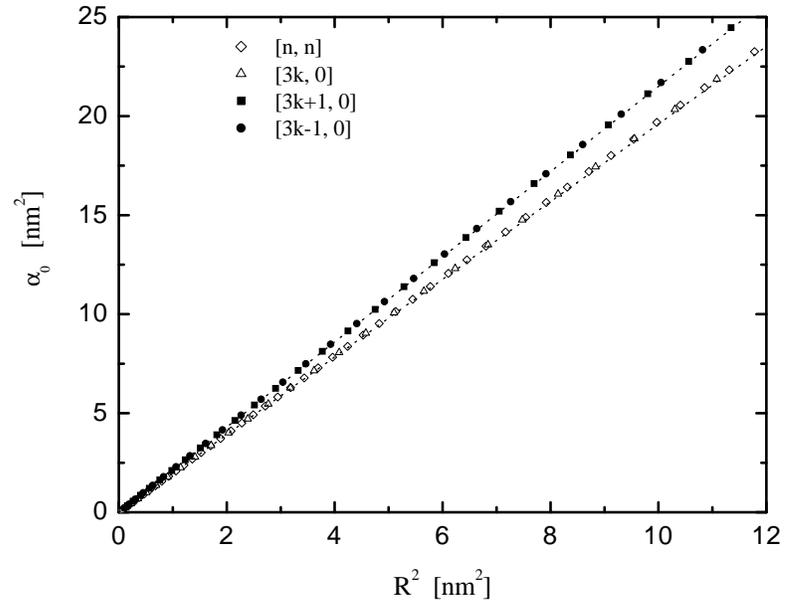}}
\caption{Plot of $\alpha_0$ vs.$R^2$ for armchair and
  zigzag tubes. The fitting lines are $\alpha_0=1.96R^2$ and
  $\alpha_0=2.15R^2$ respectively . }
\label{fig:a0}
\end{figure}

As discussed above, when the field strength is less than $ta/(eR^2)$, 
one can use wavefunctions and energies of the electron
states obtained without external perturbation.  Results of the TB calculation 
in the static limit are shown in Figure \ref{fig:a0}, where metallic and semiconducting tubes 
are compared.  The unscreened polarizability $\alpha_0$ is quadratically 
proportional to $R^2$ and is dependent on the conducting properties 
(symmetry of the tube). As the radius of NTs increases, $\alpha_0$
can be fitted by an universal expression:

\begin{equation}\label{a0_fullband}
  \alpha_0=CR^2,\quad C\approx
\left\{\begin{array}{ll}
1.96,&\mbox{metallic}\\
2.15,&\mbox{semiconducting}
\end{array}\right.
\end{equation}

\noindent
The corresponding dielectric function $\epsilon$ is

\begin{equation}\label{dielectric}
\epsilon\approx
\left\{\begin{array}{ll}
4.92,&\mbox{metallic}\\
5.30,&\mbox{semiconducting}
\end{array}\right.
\end{equation}

\noindent
which agrees very well with previous results \cite{Louie, Novikov}.  For comparison, 
we have calculated the transverse polarizability using the 
$\vec{k} \cdot \vec{p}$ scheme \cite{kp}. This approximation is equivalent 
to the linerization of the electron energy dispersion in the vicinity
of the Fermi points $\pm k_F$. The cutoff of $k_{max}=1/R$ is 
used for the calculation and only the lowest subbands are included.

With these approximations, the analytical expression for the
polarizability is

\begin{equation}\label{a0_kp}
\alpha_0=C'R^2, \quad C'=
\left\{\begin{array}{lll}
(\sqrt{2}-1)\frac{4e^2}{v_F\pi}\approx 1.41,&\mbox{metallic}\\
\beta\frac{4e^2}{v_F\pi}\approx 1.58,&\mbox{semiconducting}
\end{array}\right.
\end{equation}

\noindent
Here, $v_F=\sqrt{3}a t/2\approx5.4 \mbox{eV\AA}$ is the Fermi velocity, with $\hbar$ taken to
be unity. We also have for the integration constant, $\beta\approx 0.463$.  The value of $\alpha_0 $, 
calculated with the $\vec{k} \cdot \vec{p}$ scheme in
Eq.~(\ref{a0_kp}), is smaller than the numerical result
obtained from by the full-band calculation, Eq.~(\ref{a0_fullband}),  
but it provides a more intuitive understanding of the physical factors which affect 
the polarizability. 

The independence of the dielectric function on the geometry (NT
radius) holds only as long as one can verify the 
condition that the NT bands are half-filled.  However, we notice that some nanotubes are naturally 
doped during the growth process, in which case the Fermi level $E_F$ would shift away 
from the charge neutrality level, and similarly under conditions of charge injection or 
application of an external bias to the nanotube. 
In these cases, the polarizability depends on the occupation or depletion of new states.

Let's assume $E_F>0$ (note that the sign of $E_F$ does not affect the results) 
and define the unscreened polarizability for a metallic tube as:

\begin{eqnarray}
\alpha_0^{met}(E_F,R)&=&\alpha_0^{met}(0,R) +
\Delta\alpha_0^{met}(E_F,R)\nonumber\\
&=&\alpha_0^{met}(0,R)+\Delta\alpha_0^{met,1}(E_F,R)+\Delta\alpha_0^{met,2}(E_F,R)+\cdots
\end{eqnarray}

\noindent
where $\Delta\alpha_0^{met,i}(E_F,R)$ accounts for the contribution
due to the occupation of $i$th conduction subband. At low Fermi energy, the
$\vec{k} \cdot \vec{p}$ approximation gives:

\begin{eqnarray}\label{Da0_kp_Ef_met}
\Delta\alpha_0^{met,1}(E_F,R)&=&\frac{4e^2}{\pi v_F^3}R^4E_F^2\nonumber\\
\Delta\alpha_0^{met,2}(E_F,R)&=&\frac{8e^2}{3\pi v_F^3}R^4H[E_F-E_{b2}]\left(E_F\sqrt{E_F^2-E_{b2}^2}
-E_{b2}^2\mathrm{arcsinh}\left[\frac{\sqrt{E_F^2-E_{b2}^2}}{E_{b2}}\right]\right)
\end{eqnarray}

\noindent
Here, $H$ is the unit step function ($H[x]=1$ when $x>0$ and $0$ otherwise), and 
$E_{b2}=v_F/R$ is the energy of the bottom of the $2^{nd}$ subband.

If one considers the low energy properties of nanotubes, the $\vec{k} \cdot \vec{p}$ 
method provides a reasonable approximation. When the Fermi level is within the first
subband, the unscreened polarizability and the dielectric function for a
metallic tube of radius R can be written as:

\begin{equation}\label{a0_kp_Ef_met}
\alpha_0^{met}(E_F,R)\approx \alpha_0^{met}(0,R)+\frac{4e^2}{\pi v_F}\left(\frac{E_FR}{v_F}\right)^2R^2
\end{equation}

\begin{equation}\label{epsilon_kp_Ef_met}
\epsilon^{met}(E_F,R)\approx \epsilon^{met}(0)+ 2\frac{4e^2}{\pi v_F}\left(\frac{E_FR}{v_F}\right)^2
\end{equation}

\noindent
where the prefactor $4e^2/\pi v_F$ is the dimensionless density of states of the first subband.

As shown above, the shift of Fermi level away from the charge
neutrality level changes considerably the polarizability and, consequently, the nanotube dielectric 
function.  The dielectric screening increases with the radius of 
the tube, according to Eq.~(\ref{epsilon_kp_Ef_met}), in contrast to the screening of a neutral tube, 
which is independent of the radius. This is because of the strong dependence on $R$ found in 
the transition energy of the main terms in the expressions for $\epsilon$, $\Delta E\sim v_F/R$. 
One must conclude that the dielectric properties of a nanotube depend not only on radius 
and type of bandstructure (metal or semiconductor), but 
also on the charge carrier density in the tube \cite{Temp_effect}. This charge density 
varies with applied field if the tube is connected to electron reservoirs and charge flow 
through the tube is allowed. Thus, the calculation of the polarizability of the nanotube in a 
device environment is a complicated problem which has to be solved self-consistently.

\section{NT Bandstructure Modification in a strong field limit}\label{sec:bandstructure}

\begin{figure}[htb]
\centerline{
\includegraphics[angle=-90, width=7 in]{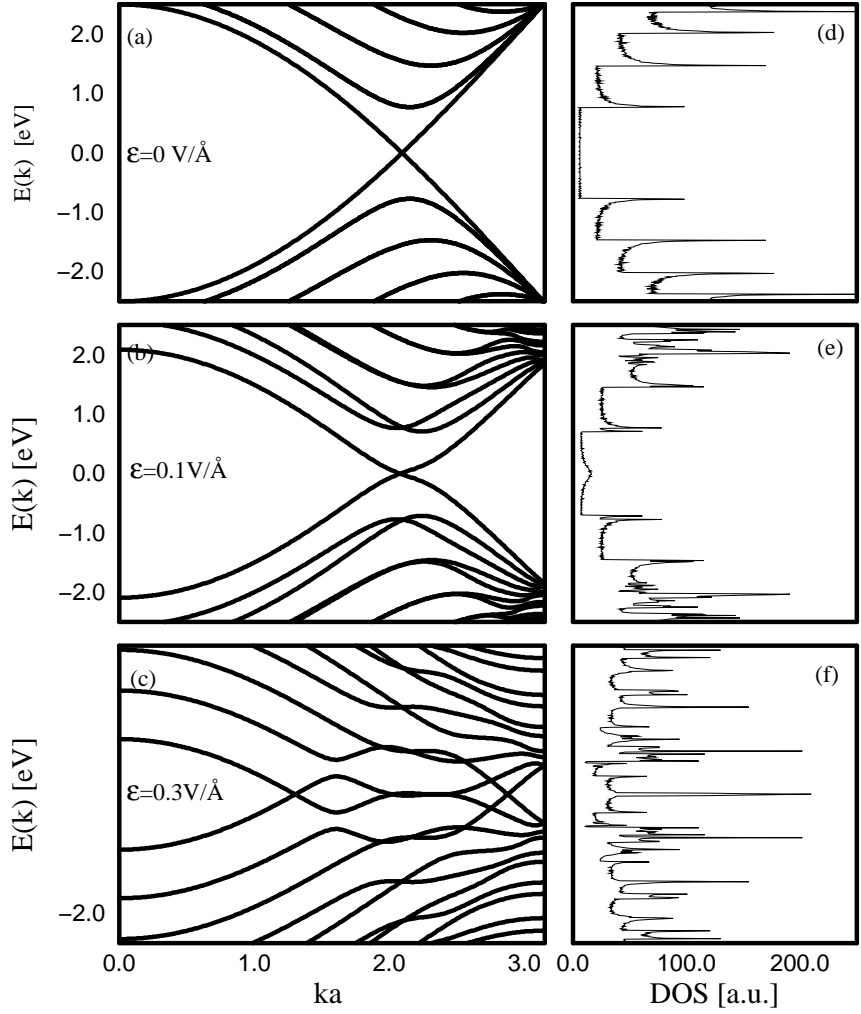}}
\caption{Band structures ((a),(b),(c)) and DOS ((d),(e),(f)) of a
  [10,10] armchair tube at various transverse electric fields: ${\cal
  E}=0,0.1,0.3\mbox{V/ \AA}$. Higher bands ($E>2.5\mbox{eV}$) are not displayed.}
\label{fig:bs_DOS_a10}
\end{figure}

When the applied field is strong enough to mix neighboring
subbands, that is $e{\cal E}R\geq ta/R$, the bandstructure of a nanotube is 
considerably modified.  We have calculated the energy bands of a [10,10] armchair
tube in electrical fields of different strength. Since we are
interested in the low energy region of nanotubes, only subbands in the energy
range $|E|\le t$ are shown in Figure ~\ref{fig:bs_DOS_a10}.
We emphasize that the two lowest subbands always cross, 
even at a very large field, although the bandstructure has been 
noticeably modified \cite{Ouyang_diff}.  At ${\cal E}=0.1 \mbox{V/\AA}$ 
(see Figure~\ref{fig:bs_DOS_a10}(b)), the Fermi points shift toward the $\Gamma$ point
($k=0$) and the two lowest subbands are flattened near the Fermi
points.  At the same time, all states which were degenerate with respect
to the magnetic number, $\pm m$, split.  The splitting becomes more obvious 
closer to the lowest subbands.  The large degeneracy at the first
Brillouin zone (FBZ) boundary $k=\pi/a$
is also lifted and bending is observed for all subbands at this
point.

As the field strength increases, the two lowest subbands show oscillatory 
bends with multiple nodes generated, while the first node moves even closer to 
$k=0$ (Figure~\ref{fig:bs_DOS_a10}(c)). For other subbands, the
splitting of $\pm m$ subbands become more significant.

This bandstructure modification is clearly seen in the density of 
states of the nanotube, as shown in Figure~\ref{fig:bs_DOS_a10}(d--f).
As the field is applied, the low energy plateau displays a bump which 
increases with field.  The enhanced DOS near $E=0$ is due to the 
flattening and bending of the two lowest subbands. On the other hand, 
the lifting of $\pm m$ degeneracy of all doublets and the bending at the 
FBZ boundary split the single VHS peaks into multiple ones. 
Although the DOS structure (Figure~\ref{fig:bs_DOS_a10}) changes considerably 
as compared to the case ${\cal E}=0$, all DOS features may be attributed 
to specific symmetry of the states.  We stress that many experimental techniques, 
ranging from Raman scattering to scanning tunneling spectroscopy use high electric
fields to probe the electronic properties of a nanotube, which may perturb
the underlying electronic structure. Our theoretical results may help to 
understand disagreements between experimental measurements and predictions for 
DOS, effective masses and the locations of VHS peaks.  
For example, we obtained different shifts of optical transition energy 
$E_{11}$ and $E_{22}$.

\begin{figure}[htb]
\centerline{
\includegraphics[angle=-90,width=5 in]{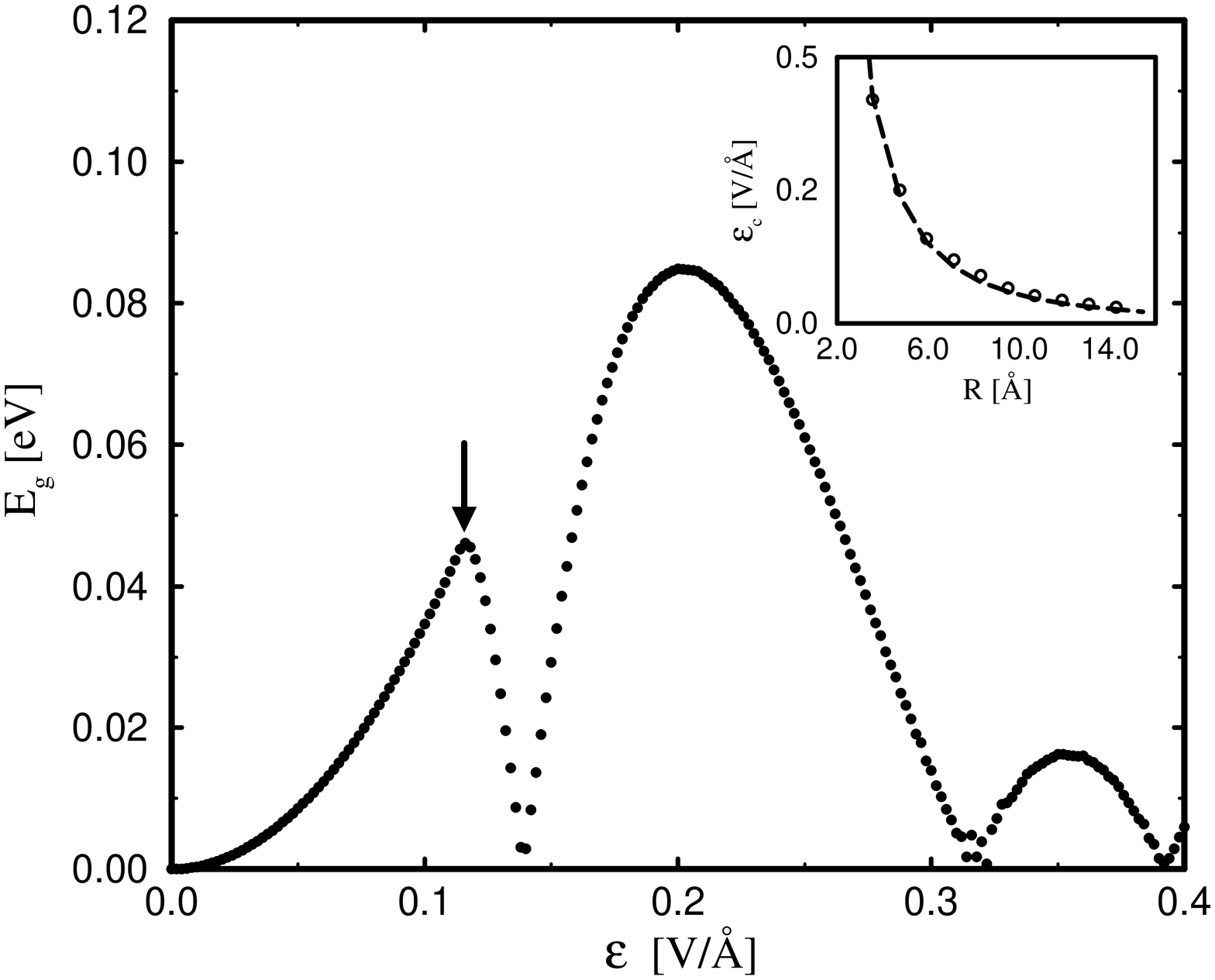}}
\caption{Bandgap variation of a [18,0] tube with increasing field
  strength. The arrow indicates the critical field strength ${\cal E}_c$
  (see text).
Inset: ${\cal E}_c$ as a function of the radius $R$. Open circles represent 
the numerical results for metallic zigzag tubes and the dashed line
is the fitting curve ${\cal E}_c=v_F/eR^2$. }
\label{fig:bgc_zmet_R}
\end{figure}

For quasi-metallic zigzag tubes, the physics is quite different. At weak fields,
a bandgap opens at the Fermi point $k=0$. When the total
field is smaller than a critical field ${\cal E}_c$, the gap
is quadratically proportional to the product of the field strength ${\cal E}$ and
the radius $R$, as shown in Figure ~\ref{fig:bgc_zmet_R}. The analytical
expression for the gap can be obtained within a degenerate perturbation theory:

\begin{equation}\label{perturbation}
E_g\sim \frac{(e{\cal E} R)^2}{6t}, \,{\cal E}<{\cal E}_c
\end{equation}

\noindent
When the field increases beyond  ${\cal E}_c$, the minimum gap 
shifts away from the original Fermi point $k=0$ and the gap starts 
to decrease. The value of the critical field depends on the nanotube 
radius (see Figure~\ref{fig:bgc_zmet_R} inset) and is fitted to be:

\begin{equation}\label{fit:Ec}
e{\cal E}_c R\sim v_F/R
\end{equation}

\noindent
Thus, the degenerate perturbation theory which is used in deriving Eq.~(\ref{perturbation}) 
is no longer valid, when the external potential exceeds the energy distance between 
neighboring subbands, $\Delta E \propto v_F/R$.

It may be interesting for electronics applications to be able to 
modulate locally the gap of the nanotube. Our study shows that for metallic 
zigzag NT one can, indeed, open the gap. However the gap cannot exceed some 
critical value beyond which a further increase of the field 
begins to close the gap.  From Eqs.~(\ref{perturbation}) and ~(\ref{fit:Ec}), 
the critical bandgap for a metallic zigzag tube $[3n,0]$ is approximately:

\begin{equation}\label{fit:Egc}
E_{gc}\sim\frac{v_F^2}{6tR^2}=\frac{\pi^2}{2n^2}t
\end{equation}

which is almost negligible for nanotubes with large radius
($n>20$).  Although larger radius NTs may seem more attractive due to the 
smaller critical field is smaller, in these structures the maximum gap will 
also be smaller. On the other hand, we notice that for very narrow NTs, the $\sigma-\pi$
mixing may results in the opening of secondary gaps, which may
prohibit using very narrow NTs for band modulations.

The transverse field effect discussed above, including gap opening or preserved 
subband crossing as well as degeneracy lifting, can be explained by 
using methods of group theory.  For a detailed description of a group
theory technique for nanotubes, we refer the reader to \cite{Vukovic}. 
For armchair and zigzag tubes, the full set of quantum numbers includes 
the longitudinal momentum $k$, the angular momentum $m$, and parities with
respect to the vertical mirror reflection $\sigma_v$ and the
horizontal mirror reflection $\sigma_h$ \cite{Vukovic}. Of the
$4n$ bands (for a $[n,n]$ or $[n,0]$ tube), only those with  $m=0,n$ have 
definite even or odd parity about $\sigma_v$ and they are non-degenerate 
(spin degeneracy is not relevant here and is not discussed). 
All the other bands are doubly degenerate at zero field with 
respect to $+m$ and $-m$ indices.

We assume that the components of the electron state on different sub-lattices 
are not mixed when the magnitude of the field satisfies 
$e{\cal E} a_{c-c}\ll t$, where $a_{c-c}$ is the nearest neighbor distance in 
the CNT. Thus, a uniform transverse electric field possesses an odd parity 
with respect to a vertical mirror plane of the nanotube.  Due to the commutation 
of a perturbed Hamiltonian $H=H_0+e{\cal E} x$ and the vertical parity 
operator $\sigma_v$, parities with respect to $\sigma_v$ remain good quantum 
numbers.  For an armchair tube, the two crossing subbands ($m=n,s=\pm 1$) 
have opposite $\sigma_v$ parities. 

In the presence of a transverse field, the degeneracy of these two subband 
at $k_F=2\pi/3a$ is lifted in the second order of perturbation theory, but 
since they have different $\sigma_v$ parities, the crossing at the new Fermi 
points is not prohibited. In contrast, for metallic zigzag tubes, electron states 
in the four lowest subbands $(m=\pm 2n/3,s=\pm1)$ do not have definite 
$\sigma_v$ parities and can be mixed by the external field in a high order 
perturbation theory by coupling to the states with $m=2n/3\pm1$.  Two pairs of 
new states with opposite parities with respect to $\sigma_v$ are the result.
According to the anti--crossing rule for NTs subbands noticed in \cite{Noncrossing},
the degeneracy at $k_F=0$ will be lifted. Similarly, the splitting of the degenerate 
subbands $\pm m$ is explained by their mixing to the other subbands by the 
transverse field. Because of the selection rule $m'=m\pm 1$, the two degenerate 
states $|\psi_m^{(0)}\rangle$ and $|\psi_{-m}^{(0)}\rangle$ can only be mixed and split 
at a high order of perturbation theory. Thus, the coupling strength between 
$|\pm m\rangle$ states depends on min$[m,n-m]$, which is a ``distance'' 
between these two states and the non-degenerate states $|0\rangle$ and $|n\rangle$.
This explains why, at low energy, the splitting is more prominent for subbands 
in armchair tubes, i.e. near  $m=n$ and almost indiscernible for subbands in 
zigzag tubes, i.e. away from $m=0,n$. 

In summary, we studied for the first time the effect of doping and/or
charge injection on the transverse polarizability of nanotubes. We
found that for metallic tubes, the polarizability grows quadratically 
with $E_F$ and scales as $R^4$ at low Fermi levels, leading to an 
enhancement of the dielectric function for the doped nanotubes. With an 
increase of the the applied field, the bandstructure is
considerably modified due to the lowering of symmetry. The zero-gap structure of
armchair tubes is always preserved while gap opening and closing
occur in metallic and semiconducting zigzag tubes respectively. Degeneracy lifting of
$\pm m$ subbands and the flattening of lowest subbands are predicted,
which changes the shape and magnitude of the peaks in the density of
states.Thus we conclude that formation of multiple valleys in the
bandstructure, enhancement of DOS at the Fermi level and engineering
of subbands with required effective mass of charge carriers are
possible with a transverse electric field. We also predict a maximum
bandgap which opens in a metalic tube and give an optimal range of nanotube
radius for bandgap engineering. 

\section*{ACKNOWLEDGMENTS}
  This work was partial supported by the National Science Foundation,
grant ITR/SY 0121616. One of us (SVR) acknowledges support through a
  CRI grant of UIUC, DoE grant DE-FG02-01ER45932, NSF grant
  ECS-0210495 and Beckman Fellowship from the Arnold and Mabel Beckman Foundation

\newpage

\newpage

\newpage


\begin{thebibliography}{9}
\bibitem{Temp_effect} Our calculation shows that the temperature
  contribution adds to the dielectric function in
  Eq.~(\ref{epsilon_kp_Ef_met}) as  $\frac{2\pi^2}{3}\left(4e^2/\pi
  v_F\right)\left(k_B TR/v_F\right)^2$.
 This correction is smaller than $10$\% at room temperature for tubes
 with a moderate radius ($R<32\mbox{\AA} $).
\end{thebibliography}
\end{document}